\documentclass[journal=ancac3,manuscript=article]{achemso}
\setkeys{acs}{etalmode=truncate,maxauthors=20}

\usepackage{graphicx}
\usepackage[capitalize]{cleveref}
\usepackage{soul}
\usepackage{xcolor}
\usepackage[utf8]{inputenc}
\usepackage{placeins}
\usepackage{hyperref}

\usepackage{bm}
\usepackage{mathtools,amssymb}
\usepackage{siunitx}
\DeclareSIUnit\sq{\ensuremath{\Box}}

\author{Yongping Liao}
\affiliation{Aalto University School of Science, Department of Applied Physics, P.O. Box 15100, FI-00076 Aalto, Finland}
\altaffiliation{Contributed equally to this work.}

\author{Kimmo Mustonen}
\affiliation{University of Vienna, Faculty of Physics, 1090 Vienna, Austria}
\email{kimmo.mustonen@univie.ac.at}
\altaffiliation{Contributed equally to this work.}

\author{Semir Tulic}
\affiliation{University of Vienna, Faculty of Physics, 1090 Vienna, Austria}

\author{Viera Skakalova}
\affiliation{University of Vienna, Faculty of Physics, 1090 Vienna, Austria}

\author{Sabbir A. Khan}
\affiliation{Niels Bohr Institute, University of Copenhagen, 2100 Copenhagen, Denmark}

\author{Patrik Laiho}
\affiliation{Aalto University School of Science, Department of Applied Physics, P.O. Box 15100, FI-00076 Aalto, Finland}

\author{Qiang Zhang}
\affiliation{Aalto University School of Science, Department of Applied Physics, P.O. Box 15100, FI-00076 Aalto, Finland}

\author{Changfeng Li}
\affiliation{Nanoscience and Advanced Materials Group, Department of Micro- and Nanosciences, Micronova, Aalto University, P.O. Box 13500, FI-00076 Aalto, Finland}

\author{Mohammad R.A. Monazam}
\affiliation{University of Vienna, Faculty of Physics, 1090 Vienna, Austria}

\author{Jani Kotakoski}
\affiliation{University of Vienna, Faculty of Physics, 1090 Vienna, Austria}

\author{Harri Lipsanen}
\affiliation{Nanoscience and Advanced Materials Group, Department of Micro- and Nanosciences, Micronova, Aalto University, P.O. Box 13500, FI-00076 Aalto, Finland}

\author{Esko I. Kauppinen}
\affiliation{Aalto University School of Science, Department of Applied Physics, P.O. Box 15100, FI-00076 Aalto, Finland}

\email{esko.kauppinen@aalto.fi}

\title{Enhanced Tunnelling in a Hybrid of Single-Walled Carbon Nanotubes and Graphene}

\begin{document}

\begin{abstract}
Transparent and conductive films (TCFs) are of great technological importance. The high transmittance, electrical conductivity and mechanical strength make single-walled carbon nanotubes (SWCNTs) a good candidate for their raw material. Despite the ballistic transport in individual SWCNTs, however, the electrical conductivity of their networks is limited by low efficiency of charge tunneling between the tube elements. Here, we demonstrate that the nanotube network sheet resistance at high optical transmittance is decreased by more than 50\% when fabricated on graphene and thus provides a comparable improvement as widely adopted gold chloride ($\mathrm{AuCl_3}$) doping. However, while Raman spectroscopy reveals substantial changes in spectral features of doped nanotubes, no similar effect is observed in presence of graphene. Instead, temperature dependent transport measurements indicate that graphene substrate reduces the tunneling barrier heights while its parallel conductivity contribution is almost negligible. Finally, we show that combining the graphene substrate and $\mathrm{AuCl_3}$ doping, the SWCNT thin films can exhibit sheet resistance as low as \SI{36}{\ohm\per\sq} at 90\% transmittance.

\end{abstract}

\flushbottom
\maketitle
\thispagestyle{empty}
\textbf{Keywords:} SWCNT, graphene, transport, conductivity, transparent and conductive films\vspace{14pt}

The electrical transport in networks of single-walled carbon nanotubes (SWCNTs) vary in a wide range of values as the structure of tubes and the morphology of networks differ. Since the modest conductivity reported in the seminal demonstrations,~\cite{wu2004transparent, hu2004percolation} the performance has gradually improved through morphological optimization~\cite{hecht2006conductivity, kaskela2010aerosol, hecht2011high, mustonen2012influence, mustonen2015gas, mustonen2015uncovering, jiang2018ultrahigh} and progress in non-covalent doping.~\cite{lee1997conductivity, choi2002spontaneous, kim2008fermi, hecht2011high} Meanwhile, as confirmed by numerous direct measurements,~\cite{fuhrer2000crossed, nirmalraj2009electrical, znidarsic2013spatially} the limiting factor in network conductivity remains to be the inefficient charge tunneling between individual tubes and thus, the central paradigm lies in their interface optimization. In this role the contacts have been bridged for example by using suitable work function metals~\cite{chen2018effect} and also more recently, with graphitized carbon welds.~\cite{jiang2018ultrahigh} The latter approach has proven particularly successful and the thin film performance (as measured by the ratio of conductance and absorbance) approaches the projected ultimate limit for SWCNT transparent electrodes.~\cite{mustonen2015uncovering, pereira2009upper} In the same spirit graphene and nanotubes have been combined into hybrid thin films,~\cite{tung2009low, gorkina2016transparent, bittolo2010electrodeposition, hong2010transparent, kholmanov2015optical, kim2009durable, li2010graphene, liu2011production, nguyen2012controlled, yang2015highly} although with performance not higher than that has been separately reported for pristine SWCNTs.~\cite{hecht2011high, kaskela2010aerosol, nasibulin2011multifunctional, mustonen2015gas} 

In this article, we present a detailed study of charge transport in SWCNT networks on a dielectric surface and on graphene; a division that has not been previously addressed in required detail. In contrast to most SWCNTs deposited through liquid phase, we have used a floating catalyst synthesis approach~\cite{liao2018direct} that does not compromise the tube cleanliness and quality by surfactant treatments. By using the same nanotube raw material, we have fabricated thin films on both a dielectric and a graphene substrate and have studied their charge transport and optical transmittance in pristine and doped states. On graphene, the SWCNT conductivity is found to be increased by a similar amount as is induced by chemical doping. Nevertheless, contrary to chemical doping, Raman spectroscopic measurements indicate no evidence of charge transfer between the nanotube and graphene layers. Instead, we establish the presence of graphene decreases the tunneling barrier heights and thus results in efficient inter-tube charge transport and hence greatly improved conductivity.

\newpage

\section*{Results and Discussion}
\begin{figure*}[!ht]
\centering
\includegraphics[width=\textwidth,keepaspectratio]{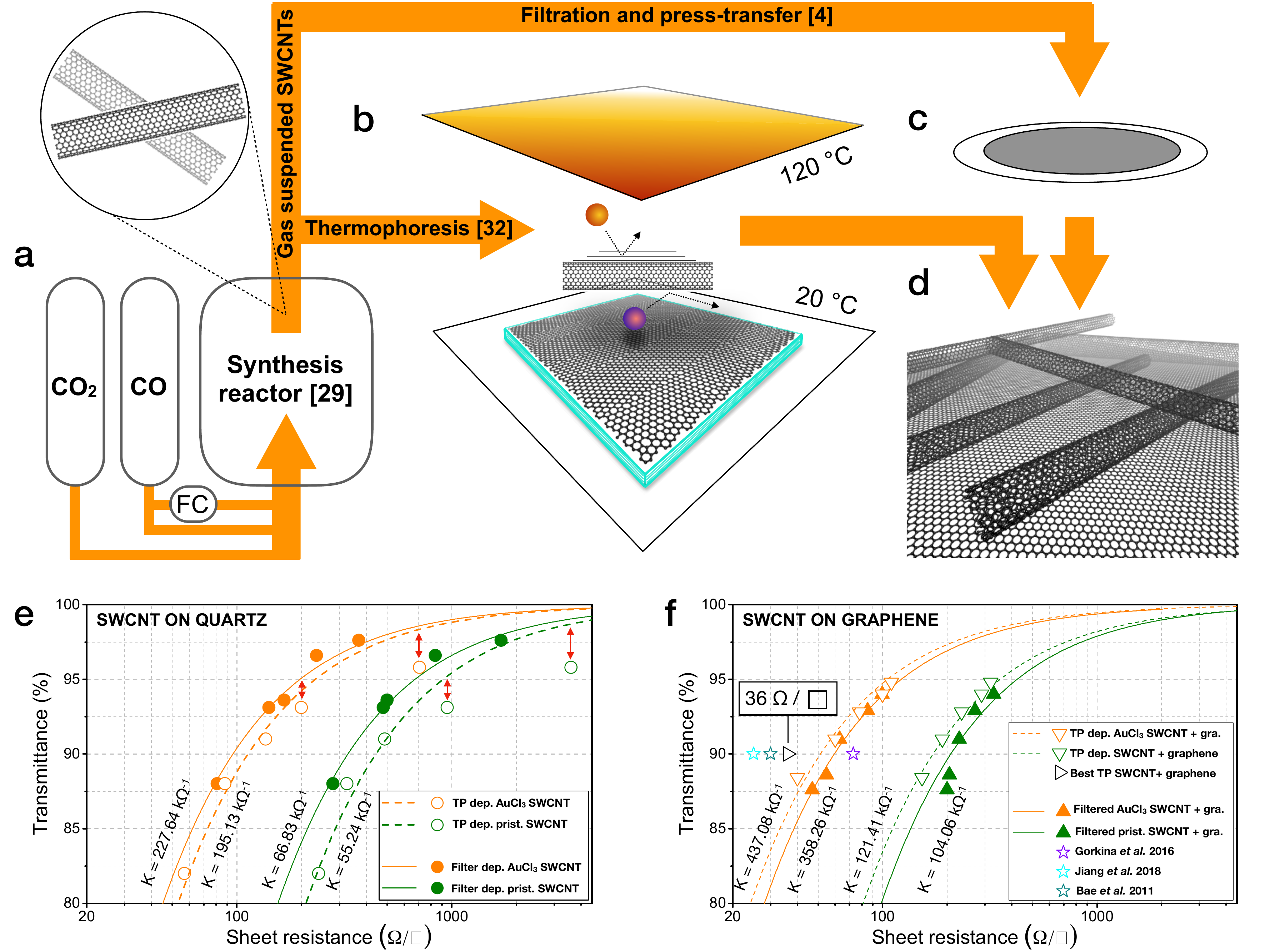}
\caption{\textbf{TCF fabrication and the optoelectronic performance.} (a) A schematic of the floating catalyst synthesis reactor.~\cite{moisala2006single, liao2018direct} FC, CO and CO$_2$ stand for the ferrocene cartridge, and carbon monoxide and dioxide gas cylinders. (b) A schematic depiction of direct SWCNT deposition using thermophoretic precipitator~\cite{laiho2017dry} (TP) on graphene electrode and (c) on a membrane filter.~\cite{kaskela2010aerosol} (d) An artistic rendition of SWCNTs interfacing with graphene. (e) The sheet resistance vs. transmittance data for SWCNTs on quartz and (f) SWCNTs on graphene. The curves are fits to Equation~\ref{quality_factor}.}
\label{fig:Figure_1}
\end{figure*}

The SWCNT raw material was grown in a vertically assembled floating catalyst reactor (see Methods and Figure~\ref{fig:Figure_1}a).~\cite{moisala2006single, liao2018direct, liao2018tuning} With the same approach, some of us have earlier demonstrated that SWCNT properties can be tuned by small changes in the composition of synthesis atmosphere.~\cite{liao2018tuning} In this work we used a composition that concurrently maximizes both the tube diameter and length, which according to earlier electron microscopy experiments correspond to 1.9$\pm$0.5 nm and 7.5$\pm$5.6 $\mathrm{\mu}$m, respectively.~\cite{liao2018direct} For optical and electrical characterization, the nanotube films were either accumulated directly on the target substrate by using a thermophoretic precipitator (TP)~\cite{laiho2017dry} or, for reference purpose, by vacuum filtration and press-transfer (see Figure~\ref{fig:Figure_1}b-c).~\cite{kaskela2010aerosol}

The sheet conductance ($\mathrm{\sigma}$) and optical absorbance (A) of a uniform SWCNT film are directly related. This constant of proportionality can be understood as the quality factor (K) of the carbon nanotube raw material. Written using the Beer-Lambert law describing the attenuation of light in continuum media and sheet resistance ($\mathrm{R_s}$) yields
\begin{equation}\label{quality_factor}
	\mathrm{K = {\sigma}\times{A^{-1}} = {[R_{s}\times\log_{10}(T)]}^{-1}},
\end{equation}
where the latter equality is an alternative expression using optical transmittance ($\mathrm{T}$). Thus, when the network's density is far above the percolation threshold,~\cite{pike1974percolation} $\mathrm{R_s}$ and T have a log-linear relation. The measured values for different SWCNT film thicknesses gathered for our experiments are shown in Figure~\ref{fig:Figure_1}e, and their $\mathrm{R_s}$ vary in a wide range from $\sim$\SI{170}{\ohm\per\sq} up to $\sim$\SI{3.5}{\kohm\per\sq}. The trendlines visible in the plots are fitted according to Equation~\ref{quality_factor}.

SWCNT films made by the two deposition techniques exhibit slightly different sheet resistance characteristics. At for example 90\% transmittance the $\mathrm{R_s}$ of TP deposited (as estimated from the trendlines) equals to $\sim$\SI{450}{\ohm\per\sq}, whereas the corresponding value for filtrated networks is $\sim$\SI{330}{\ohm\per\sq}. The sheet resistance scaling as a function of optical density (or films thickness) is also very different. At low density (T$\geq$92\%), the TP deposited nanotubes show, reproducibly $\mathrm{R_s}$ that is higher than predicted by the bulk model in Equation~\ref{quality_factor} (see the highlighted data in Figure~\ref{fig:Figure_1}e). Such is, however, not the case for filtrated networks as is also corroborated by our earlier similar experiments.~\cite{kaskela2010aerosol, mustonen2015uncovering, nasibulin2011multifunctional} Since all TCFs were fabricated from virtually identical SWCNTs, these qualitative differences must emerge from the organization of the individual tubes and their interconnections, which we will discuss later. The addition of graphene layer in between the quartz substrate and the nanotube network decreased the $\mathrm{R_s}$ substantially (Figure~\ref{fig:Figure_1}f). The largest change was observed for TP deposited SWCNTs, dropping from $\sim$\SI{450}{\ohm\per\sq} to $\sim$\SI{180}{\ohm\per\sq} and thus totalling 60\%. This is particularly interesting, since the $\mathrm{R_s}$ of substrate-supported graphene is much higher, typically in the range of \SIrange{700}{1000}{\ohm\per\sq}. Meanwhile the decrease in filtered (and press-transferred) networks was a more moderate 35\%, equalling a drop from $\sim$\SI{330}{\ohm\per\sq} to $\sim$\SI{215}{\ohm\per\sq}.

The films were next treated with 16 mM gold chloride in acetonitrile solution ($\mathrm{AuCl_3}$, see Methods), further decreasing the $\mathrm{R_s}$. The lowest $\mathrm{R_s}$ was achieved with the TP SWCNTs on graphene, yielding on average $\sim$\SI{50}{\ohm\per\sq} at 90\% transparency. Also, the doping seems to improve the performance by the same factor regardless of the initial sheet resistance. Thus for example the filtered SWCNTs on graphene only reached $\sim$\SI{65}{\ohm\per\sq} but on quartz they still performed better than the TP SWCNTs ($\sim$\SI{95}{\ohm\per\sq} vs. $\sim$\SI{115}{\ohm\per\sq}). The best TP deposited SWCNTs reached a value as low as $\sim$\SI{36}{\ohm\per\sq}, which is among the lowest reported for any carbon based TCFs.~\cite{gorkina2016transparent, jiang2018ultrahigh, bae2010roll}

One could argue whether the observed improvement in presence of graphene is emerging from current distributing over the two parallel conduction layers (graphene and nanotubes). As intriguing as this idea is, it fails to provide even remotely correct predictions. Calculating for example the combined $\mathrm{R_s}$ using $\sim$\SI{750}{\ohm\per\sq} (the resistance of TP deposited SWCNTs at T=$\sim$92.5\%, Figure~\ref{fig:Figure_1}e) and an optimistic value of $\sim$\SI{650}{\ohm\per\sq} for graphene yields $\sim$\SI{350}{\ohm\per\sq}. This result is roughly twice as high as the measured $\sim$\SI{180}{\ohm\per\sq}. For filtered networks the discrepancies are clearly smaller (10-20\%), indicating the layers are much less interconnected and are better described by the parallel approximation.

We next turned our attention to the mechanisms that could explain the observed improvement. From earlier contributions we know that charge tunneling efficiency between individual tubes can be visualized by temperature dependence of conductance. The mechanism is well understood within the framework of so called fluctuation-assisted tunneling (FAT) model,~\cite{sheng1980fluctuation} often amended with an additional term describing the phonon backscattering.~\cite{pietronero1983ideal, kaiser2001electronic} For sheet resistance it can be written as:
\begin{equation}\label{R_T}
	\mathrm{R(T) = A\times\exp(-\frac{T_m}{T})+B\times\exp(\frac{T_b}{T_s+T})},
\end{equation}
where the geometric factors A and B can be taken as constants. Here the first term reflects the backscattering by lattice vibrations with a characteristic phonon energy $\mathrm{k_B T_m}$ ($\mathrm{k_B}$ being the Boltzmann constant) and the second term fluctuation-assisted tunneling through the energy barriers, $\mathrm{k_B T_b}$, dividing the metallic regions. The parameter $\mathrm{T_s}$ is the tunneling efficiency near T=0 K where conductance reaches a value $\mathrm{(B\times\exp[T_b/T_s]})^{-1}$.

\begin{figure}[ht!]
\centering
\includegraphics[width=\textwidth,keepaspectratio]{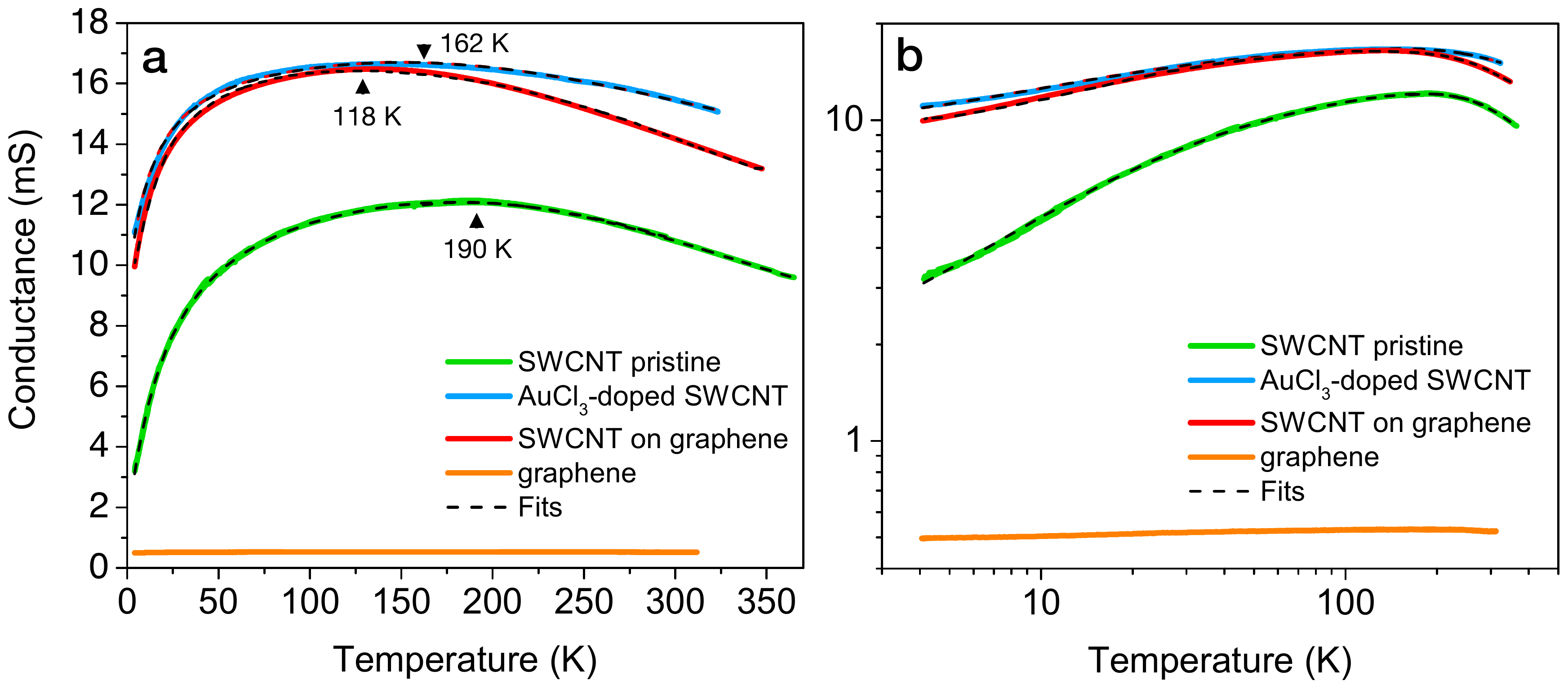}
\caption{\textbf{Temperature dependent conductance.} The G(T) plots for TP deposited SWCNTs on $\mathrm{SiO_2}$ (green), on graphene (red), doped with $\mathrm{AuCl_3}$ (blue) and for graphene (orange) with linear axis in (a) and in (b) log-log.}
\label{fig:Figure_4}
\end{figure}

Considering the anomalously low $\mathrm{R_s}$ was pronounced in the graphene-supported TP SWCNTs, we studied their transport in a liquid helium cryostat (Methods). The experimental results are presented in Figure~\ref{fig:Figure_4}, including pristine SWCNTs supported on $\mathrm{SiO_2}$ (green curve), $\mathrm{AuCl_3}$-doped SWCNTs on $\mathrm{SiO_2}$ (blue curve), graphene (orange curve), and SWCNTs on graphene (red curve). Generally the following is observed: \textbf{I.} The conductivity of the thin films containing nanotubes soar as a function of temperature until a maximum is reached, again decreasing at higher temperatures. \textbf{II.} The monolayer graphene's G(T) curve (and thus the absolute conductivity) is an order of magnitude below the 90\% transparency SWCNT films on $\mathrm{SiO_2}$. \textbf{III.} Placing SWCNTs on graphene and doping them with $\mathrm{AuCl_3}$ has a similar impact on the conductivity. \textbf{IV.} The rate of decrease of G(T) at high temperatures is steeper for SWCNTs on graphene than that of the doped nanotube films. Since in the cryostat instead of resistance we measured the sheet conductance, to fit the measurement results we also need to rewrite the FAT model as:
\begin{equation}\label{G_T}
	\mathrm{G(T) = G_0+G_1\times[\exp(-\frac{T_m}{T})+exp(\frac{T_b}{T_s+T})]^{-1}},
\end{equation}
where $\mathrm{G_0}$ and $\mathrm{G_1}$ are constants in the units of [S]. The parameters are listed in Table 1.

\begin{table}
\caption{\label{tab:table1} The parameters extracted by fitting Equation~\ref{G_T} to data in Figure~\ref{fig:Figure_4}.}
\begin{tabular}{lcccccc}
 & $G_\text{0}, S$ & $G_\text{1}, S$ & $T_\text{m}$, K & $E_\text{m}$, eV & $T_\text{b}$, K & $T_\text{s}$, K\\
\hline
Graphene & 0.0002 & 0.0004 & 1078 & 0.093 & 2.10 & 14.85\\
SWCNTs & -0.0411 & 0.0549 & 970 & 0.084 & 5.36 & 20.72 \\
SWCNTs + Graphene & -0.0196 & 0.0372 & 726 & 0.063 & 3.81 & 12.48 \\
SWCNTs + $\mathrm{AuCl_3}$ & -0.0104 & 0.0279 & 806 & 0.069 & 4.17 & 11.34 \\
\end{tabular}
\end{table}

\begin{figure*}[!ht]
\includegraphics[width=1\textwidth,keepaspectratio]{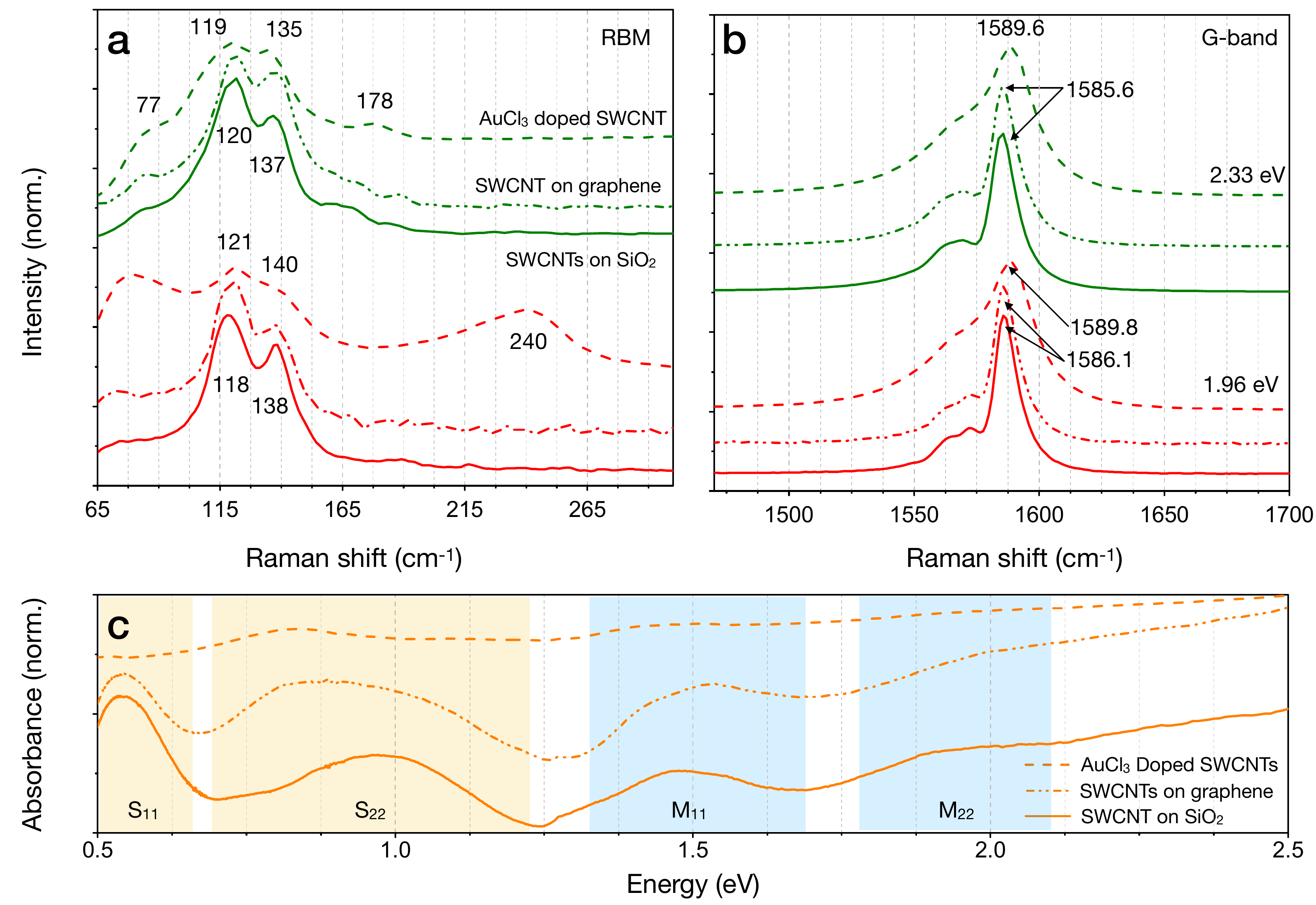}
\caption{\textbf{Spectroscopic characterization.} (a) Raman radial breathing modes of SWCNT networks on graphene, on $\mathrm{SiO_2}$ and those treated with $\mathrm{AuCl_3}$, and (b) the corresponding G-band modes (2.33 and 1.96 eV laser lines). (c) The optical absorption spectra (OAS) measured on quartz. All samples were deposited using TP.} 
\label{fig:Figure_3}
\end{figure*}

The exponentials in Equations~\ref{R_T} and~\ref{G_T} can be understood as a trade-off between the temperature-assisted charge tunneling through energy barriers that separate the metallic regions and the phonon scattering. The fitted values of $\mathrm{T_m}$ in our thin films span from 700 K to 1100 K (60-100 meV) and correspond to acoustic-phonons also visible as radial breathing modes (RBMs) in Raman spectra (see Figure~\ref{fig:Figure_3}b). The lowest value, $\mathrm{T_m}$=726 K, was observed for SWCNTs on graphene and thus reflects the lowest phonon energy with the maximum of G(T) at 118~K. Meanwhile the $\mathrm{T_m}$ for doped SWCNTs on $\mathrm{SiO_2}$ was 806 K with G(T) maximum at 162 K and for pristine SWCNTs 970 K with G(T) maximum at 190 K. Thus the phonon energies are affected both by addition of a graphene layer and by chemical doping. The tunneling barrier heights corresponding to $\mathrm{k_B T_b}$ are also affected, being $\sim$29\% lower for SWCNTs on graphene than on $\mathrm{SiO_2}$ (see $T_\text{b}$ in Table~\ref{tab:table1}). A very similar effect, however, was also observed upon chemical doping, evoking a question whether graphene also possibly acts on SWCNTs as a dopant.

This possibility can be quickly ruled out by Raman spectroscopic measurements with a G-band mode ($\sim$1580 cm$^{-1}$) blue-shift expected upon both donor or acceptor doping. No such shift, however, could be detected upon deposition on graphene (Figures~\ref{fig:Figure_3}a-b). We did, however, observe a tiny broadening of the RBMs, which could indicate stronger van der Waals (vdW) interaction with the graphene substrate. Meanwhile, the spectrum of $\mathrm{AuCl_3}$ doped SWCNTs (with a similar effect on the conductivity, see Figure~\ref{fig:Figure_4}) was clearly shifted by 4 cm$^{-1}$ and the RBMs appear to be completely changed due to a change in resonant conditions. Looking at the optical absorption spectra (OAS, Figure~\ref{fig:Figure_3}c) we observe a minor suppression of the first semiconducting transition peak ($\mathrm{S_{11}}$) on graphene. While this could indicate a mild charge transfer, it is certainly not comparable to suppression of all transitions (metallic and semiconducting) upon chemical doping. Although the transition peaks are unquestionably broader on graphene, this can be attributed to perturbations in the exciton lifetime resulting from dielectric screening~\cite{wang2006interactions} and thus supports the postulation of stronger vdW interaction in presence of graphene.

\begin{figure}[ht!]
\centering
\includegraphics[width=\textwidth,keepaspectratio]{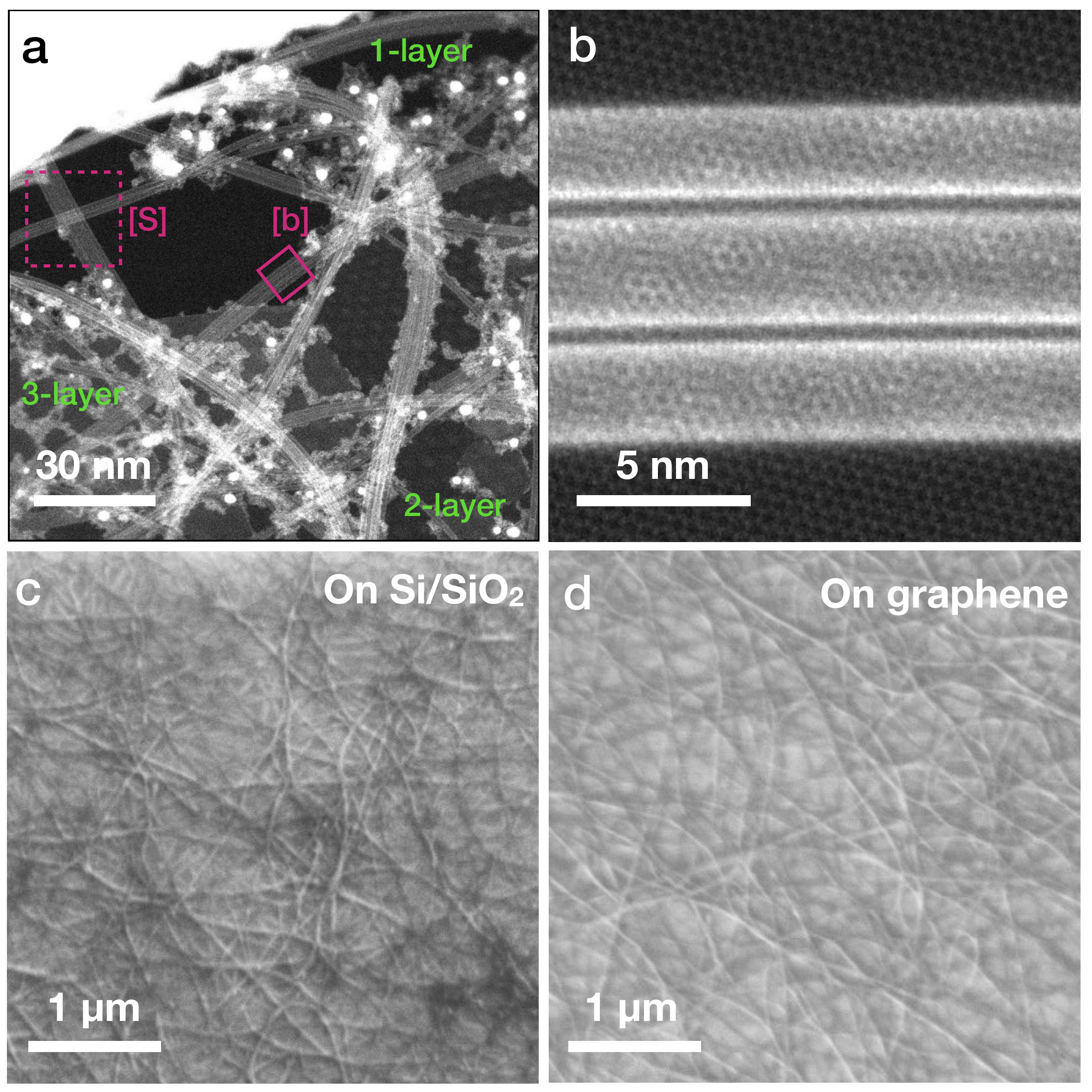}
\caption{\textbf{Thin film morphology} (a) A STEM/MAADF overview of a bundled SWCNT film on graphene. The deposition was carrier out using TP. (b) An atomically resolved close-up of a bundle firmly in contact with graphene. (c) A SEM micrograph of a thermophoretically deposited SWCNT network on $\mathrm{SiO_2}$ and (d) on graphene.}
\label{fig:Figure_2}
\end{figure}

Interfacing of SWCNTs and graphene can be studied using atomically resolved scanning transmission electron microscopy (STEM, see Methods).~\cite{mustonen2018atomic} We first exposed the sample surfaces for observations by applying laser cleaning in the microscope column,~\cite{tripathi2017cleaning} using a 10 ms pulse length with a total energy of 60 mJ. Our earlier observations revealed that when the thermodynamic constraints allow, thermophoretically deposited nanotubes form hundreds of nanometers long preferentially stacked interfaces with graphene. Our observations here suggest that regardless of higher rigidity of nanotube bundles,~\cite{liew2005buckling} similar to individual tubes, they become completely in contact with the underlying graphene substrate. Figure~\ref{fig:Figure_2}a shows an example field of view acquired by medium-angle annular dark-field detector (MAADF) with several bundles crossing on graphene, all sharing the common focus and thus, the z-height.~\cite{mustonen2018atomic} This is even more evident in the atomically resolved closeups shown in Figure~\ref{fig:Figure_2}b and Supplementary Figure~\ref{fig:SIfig2}. Now, assuming that a minor alignment of interfaces is also possible at room temperature during the nanotube deposition~\cite{mustonen2018atomic} and noting that the measurements conducted by Paulson \textit{et al.} show that this considerably improves the charge transport through the interface,~\cite{paulson2000tunable} the conductivity could also drastically improve. Indeed, if we now turn our attention to the scanning electron microscopy (SEM, see Methods) images acquired from the nanotube samples on graphene and $\mathrm{SiO_2}$ in Figures~\ref{fig:Figure_2}c-d, we observe a clear difference in their apparent contrast. As pointed out by the earlier authors,~\cite{homma2004mechanism, loos2005visualization} this kind of effect can well emerge when nanotubes are poorly interconnected, as for example when disconnected from their neighbours on a dielectric surface.

Finally, Sun \textit{et al.} concluded that the transconductance of SWCNT networks is sensitive to the inter-tube contact morphology.~\cite{sun2011flexible} Their observation was that so called tube-tube Y-junctions, which our elongated graphene/SWCNT interfaces superficially resemble, were generally more conductive than simple point contacts (X-junctions). This could explain the low barriers and low $\mathrm{R_s}$ of TP deposited SWCNTs on graphene, as they appear to be well interconnected with and through the graphene substrate. Further on, this can also explain why TP deposited SWCNTs on quartz exhibited an anomalously high $\mathrm{R_s}$ at low densities. As evident from for example Figure~\ref{fig:SIfig1}a, a low density TP deposited film is completely dominated by X-junctions and would thus be poorly interconnected without the presence of graphene. In contrast, Sun \textit{et al.} observed networks with a very similar density fabricated by filter transfer, yet they had mainly Y-junctions in their experiments.~\cite{sun2011flexible} Although they did not specifically study the mechanism of junction formation, we believe that the prominence of Y-junctions can emerge in the presence of surface roughness of the filter, providing a greater degree of freedom for the nanotubes to mutually align. In the same manner, the Y-junctions also appear in thicker TP samples (Figure~\ref{fig:SIfig1}b), indicating that the surface roughness does indeed play an important role in the formation of Y-junctions.

\section*{Conclusions}
To summarize, we have studied the mechanism of charge transport in SWCNT networks on graphene and on a dielectric substrate in their pristine and doped state. The observations show that on graphene the conductivity of nanotube networks is increased by a similar amount as is induced by chemical doping. The Raman spectroscopic measurements, however, reveal no substantial charge transfer between the graphene and nanotube layers. Instead, as probed by temperature dependent conductivity measurements, the graphene support acts as a coupling layer between the individual tubes reducing the tunneling barrier heights. This modification is responsible of enhanced interconnectivity within the binary film that, when used in combination with chemical doping, in our experiments produced sheet resistance as low as \SI{36}{\ohm\per\sq} at 90\% transmittance.

\section*{Methods}
\subsection*{Thin film fabrication}
The SWCNTs were synthesized in a vertical flow floating reactor~\cite{liao2018direct, moisala2006single} fed with a total of 400 cm$^3$ min$^{-1}$ carbon monoxide (CO) and 2 cm$^3$ min$^{-1}$ of carbon dioxide (CO$_2$). Of the CO 50 cm$^3$ min$^{-1}$ was additionally passed through a ferrocene cartridge (see FC in Figure~\ref{fig:Figure_1}a) and the mixture passed to the reactor set at 850$^{\circ}$C by using a water-cooled injector probe.~\cite{moisala2006single} At the tip of the injector SWCNTs were nucleated on the forming iron nanoparticles and grown while traversing through the hot zone. For deposition, we used either a thermophoretic precipitator (TP)~\cite{laiho2017dry} or vacuum filtration~\cite{kaskela2010aerosol} and quartz windows and silicon dioxide (SiO$_2$) were used as substrates. The TP consisted of a pair of parallel metal plates kept at $\sim$100 K temperature difference and $\sim$0.5 mm apart, thus resulting in a 20~000~$Km^{-1}$ temperature gradient. This gradient gives arise for thermophoretic force which was then used to drive the floating nanotubes on the substrate placed on the cold surface. The graphene for the experiments was acquired from Graphenea Inc. Finally, the chemical doping was done by dropcasting 16 mM gold chloride solution in acetonitrile on the samples and allowing several minutes of reaction time. Finally, the samples were flushed with pure acetonitrile and left to dry.

\subsection*{Sheet resistance measurements}
Sheet resistances were measured by using a Jandel Engineering Ltd. General Purpose 4-point probe system with a RM3000 test unit for resistance readout. The probe head pin layout is a linear array of tungsten needles with a spatial separation of 1 mm.

\subsection*{Spectroscopic measurements}
Optical absorption spectra were acquired with an Agilent Cary 5000 spectrophotometer. The samples were supported on 1 mm thick optics grade quartz windows and their contribution was omitted by placing a clean substrate on the reference beamline.

Raman experiments were conducted using a Witech Alpha300 R combined confocal Raman spectroscope and atomic force microscope using 532 nm diode and 633 nm helium-neon laser sources. The nominal power at the sample was set to $\sim$0.5 mW with a spot size of $\sim$500 nm.

\subsection*{Temperature dependent conductance measurements} 
These measurements were conducted on networks thermophoretically deposited on square SiO$_2$ substrates (size 4~mm $\times$ 4~mm) with nominal transmittance of $\sim$80\% (estimated from the collection time). The four symmetrically placed contact electrodes were fabricated by evaporating gold through a slit mask and manually wire-bonded to a Kyoreca chip carrier.

The measurement apparatus consisted of a liquid helium tank with custom-built vertically movable sample arm including a thermocouple for temperature readout and a Keithley 2635B sourcemeter for 4-point conductivity measurements. Before slowly immersing the sample arm into liquid helium (He), the volume was evacuated to a pressure of 10$^{-3}$ mbar and finally filled with He gas. The cooling rate was kept at $\sim$10 K~min$^{-1}$.

\subsection*{Scanning transmission electron microscopy (STEM)}
The electron microscopic imaging was done in an aberration-corrected Nion UltraSTEM 100 operated with a 60-keV primary beam energy, with the sample in ultrahigh vacuum (5$\times$10$^{-10}$ mbar). The angular range for the medium-angle annular dark-field (MAADF) detector was 60--200 mrad. The samples were cleaned with a 6 W continuous wave laser directly attached to the microscope column.~\cite{tripathi2017cleaning}

\subsection*{Scanning electron microscopy (SEM)}
The images were acquired by using a Zeiss Supra 55 VP analytical SEM with beam energy of 5 kV using the in-lens secondary electron detector.

\begin{acknowledgement}
This work was supported by the Academy of Finland \textit{via} projects 286546-DEMEC and 292600-SUPER, by TEKES Finland \textit{via} projects 3303/31/2015 (CNT-PV) and 1882/31/2016 (FEDOC), and the Aalto Energy Efficiency (AEF) Research Program through the MOPPI project. The authors also thank the funding from the Austrian Science Fund (FWF) under project nos. P 25721-N20, I1283-N20, P 28322-N36 and I3181-N36 and K.M. acknowledges support from the Finnish Foundations’ Post Doc Pool. J.K. acknowledges funding from Wiener Wissenschafts- Forschungs- und Technologiefonds through project MA14-009.
\end{acknowledgement}\newline

Supporting Information Available. STEM images of SWCNT networks on graphene with varying thickness \textit{via} the Internet at http://pubs.acs.org.

\bibliography{sample}

\newpage
\section*{Supporting Information}

\renewcommand{\thefigure}{S\arabic{figure}}
\setcounter{figure}{0}

\begin{figure*}[ht!]
\includegraphics[width=8.5cm,keepaspectratio]{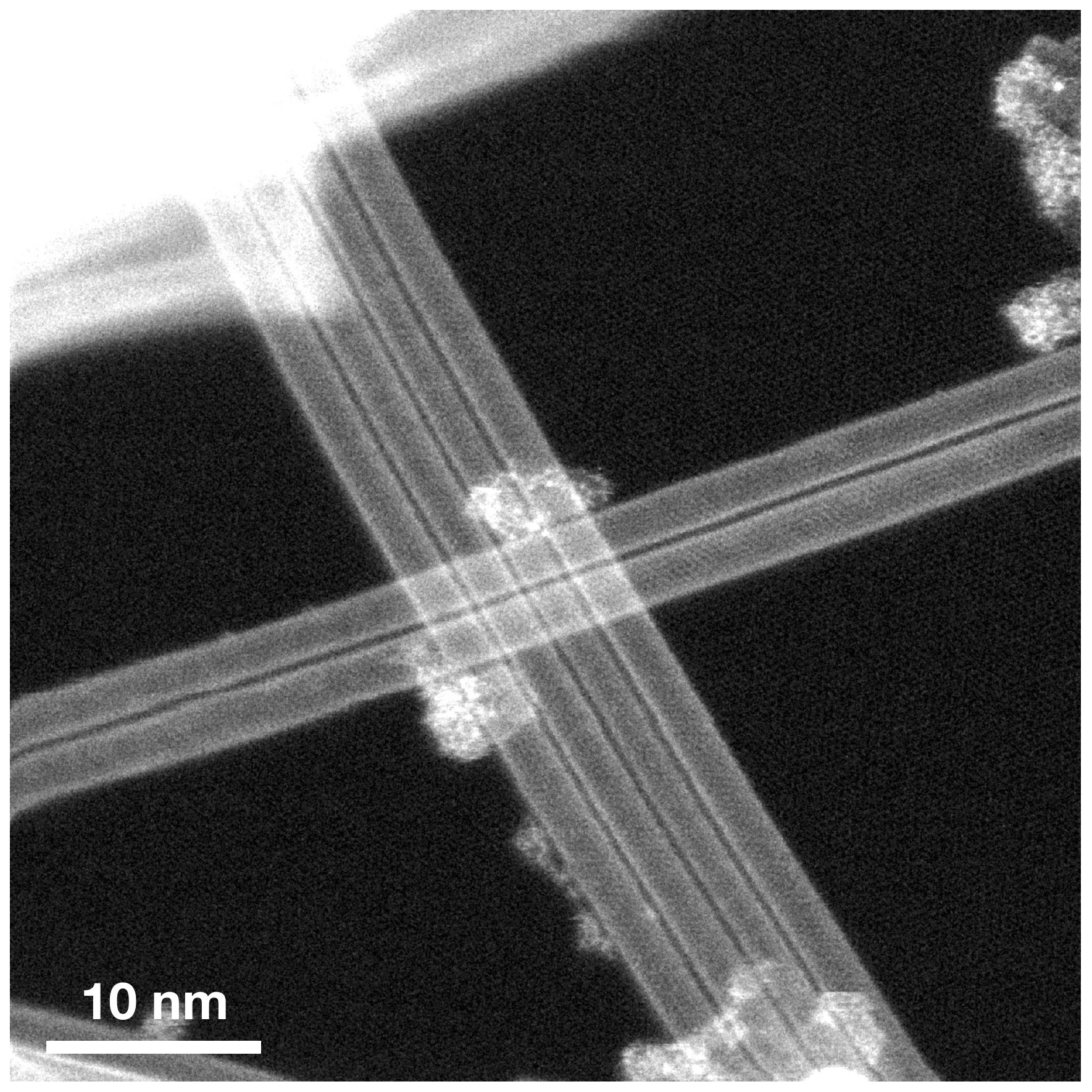}
\caption{STEM/MAADF closeup of two bundles forming an X-junction in Figure~\ref{fig:Figure_4}.\label{fig:SIfig2}}
\end{figure*}

\begin{figure*}[ht!]
\includegraphics[width=15.5cm,keepaspectratio]{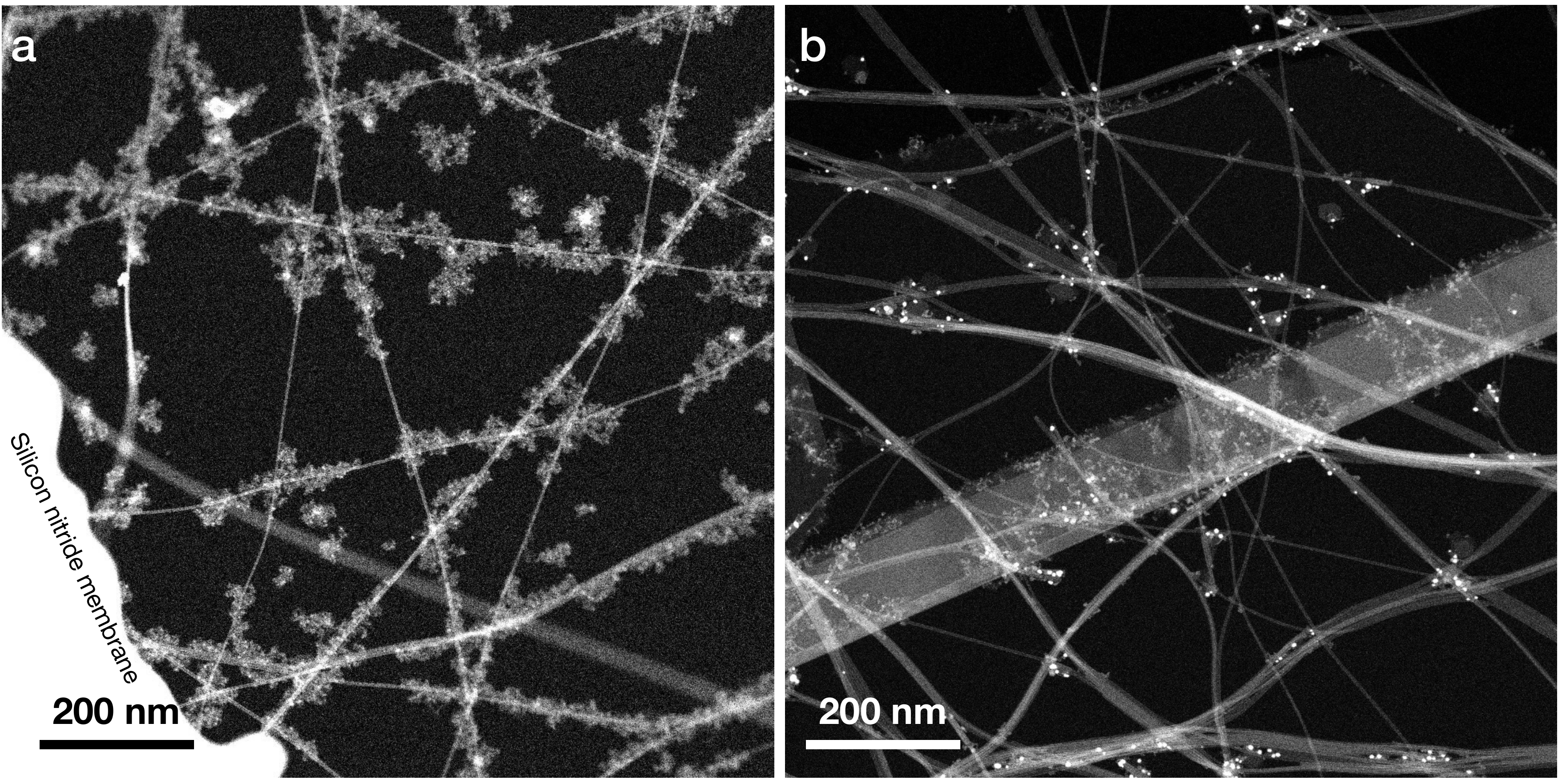}
\caption{STEM/MAADF images of different SWCNT layer thicknesses on laser cleaned graphene.~\cite{tripathi2017cleaning} Note how the morphology qualitatively changes from one dominated by X-junctions to Y-junctions when the deposition time is increases from 2 minutes in (a) to 10 minutes in (b).\label{fig:SIfig1}}
\end{figure*}

\end{document}